\documentclass[a4paper,11pt]{article}

\usepackage{jcappub}
\usepackage{slashed}
\usepackage{bm}
\usepackage{latexsym,amssymb,amsmath,float,url,mathrsfs}
\usepackage{tikz}
\usetikzlibrary{tikzmark}
\usepackage{latexsym}
\usepackage{epstopdf}
\usepackage{amsfonts}
\usepackage{amsmath}
\usepackage{amssymb}
\usepackage{comment}
\usepackage{subfigure}
\usepackage{natbib}
\usepackage{hyperref}
\usepackage{pifont}
\usepackage{blindtext}
\usepackage{adjustbox}
\usepackage{multirow}
\usepackage{tabularx}
\usepackage{tikz}
\usetikzlibrary{tikzmark}
\usepackage{amssymb}
\usepackage{bbold}
\usepackage{blkarray}
\usepackage{graphicx}
\usepackage{amsfonts}
\usepackage{soul}
\usepackage{amssymb}
\usepackage{amsmath}
\usepackage{cancel}
\usepackage{tcolorbox}

\usepackage{graphics,appendix,afterpage,makecell} 

\definecolor{oucrimsonred}{rgb}{0.6, 0.0, 0.0}
\definecolor{persianblue}{rgb}{0.11, 0.22, 0.73}
\definecolor{forestgreen}{rgb}{0.13,0.35,0.13}
\definecolor{lightgray}{rgb}{0.83, 0.83, 0.83}
 \hypersetup{colorlinks, citecolor=oucrimsonred, linkcolor=black, urlcolor=oucrimsonred}
\definecolor{cornellred}{rgb}{0.7, 0.11, 0.11}
\definecolor{navyblue}{rgb}{0.0, 0.0, 0.5}
\definecolor{amethyst}{rgb}{0.6, 0.4, 0.8}
\definecolor{yellow}{rgb}{1.0, 1.0, 0.0}
\definecolor{firebrick}{rgb}{0.7, 0.13, 0.13}
\definecolor{tangerineyellow}{rgb}{1.0, 0.8, 0.0}
\definecolor{deepfuchsia}{rgb}{0.76, 0.33, 0.76}
\definecolor{amber}{rgb}{1.0, 0.75, 0.0}
\definecolor{VioletRed4}{rgb}{0.55, 0.13, .32}
\definecolor{indiagreen}{rgb}{0.07, 0.53, 0.03}
\definecolor{VioletRed4}{rgb}{0.55, 0.13, .32}
\newcommand{\be}{\begin{equation}}
\newcommand{\ee}{\end{equation}}
\newcommand{\bea}{\begin{equation} \begin{aligned}}
\newcommand{\eea}{\end{aligned} \end{equation}}

\definecolor{oucrimsonred}{rgb}{0.6, 0.0, 0.0}
\newcommand\vertarrowbox[3][6ex]{%
  \begin{array}[t]{@{}c@{}} #2 \\
  \left\uparrow\vcenter{\hrule height #1}\right.\kern-\nulldelimiterspace\\
  \makebox[0pt]{\scriptsize#3}
  \end{array}%
}
\hypersetup{
     colorlinks   = true,
     citecolor    = violet,
     urlcolor     = violet,
     linkcolor    = violet}

\definecolor{verdechiaro}{rgb}{0.6,1,0.6}
\definecolor{giallochiaro}{rgb}{1,1,0.6}
\definecolor{bluscuro}{rgb}{0.15, 0.2, 0.9}
\definecolor{verdes}{rgb}{0.1, 0.5, 0.1}%
\definecolor{tangerineyellow}{rgb}{1.0, 0.8, 0.0}

\definecolor{americanrose}{rgb}{1.0, 0.01, 0.24}
\definecolor{cobalt}{rgb}{0.0, 0.28, 0.67}
\definecolor{brandeisblue}{rgb}{0.0, 0.44, 1.0}
\definecolor{mycolor}{rgb}{0.0, 0.0, 0.5}
\definecolor{oxfordblue}{rgb}{0.0, 0.13, 0.28}
\definecolor{azure}{rgb}{0.0, 0.5, 1.0}
\definecolor{turquoiseblue}{rgb}{0.0, 1.0, 0.94}
\newtcolorbox{mynewbox}[1]{colback=white!5!white,colframe=azure!75!black,fonttitle=\bfseries,title=#1}
\newtcolorbox{mybox}{colback=mycolor!5!white,colframe=azure!75!black}
\newtcolorbox{mynamedbox}[1]{colback=mycolor!5!white,colframe=azure!75!black,title=#1}
\definecolor{venetianred}{rgb}{0.78, 0.03, 0.08}
\newtcolorbox{mynamedbox1}[1]{colback=venetianred!5!white,colframe=venetianred!80!black,title=#1}
\newtcolorbox{mynamedbox2}[1]{colback=azure!5!white,colframe=azure!80!black,title=#1}

\definecolor{verdes}{rgb}{0.1, 0.5, 0.1}%
\definecolor{cornellred}{rgb}{0.7, 0.11, 0.11}

\definecolor{VioletRed4}{rgb}{0.55, 0.13, .32}

\definecolor{rossocorsa}{rgb}{0.83, 0.0, 0.0}

\title{The Nonlinear Tails in Black Hole Ringdown:
the Scattering Perspective  }

\author[a]{A.~Ianniccari,}
\author[a,b]{L.~Lo Bianco,}
\author[a]{A.~Riotto}

\affiliation[a]{Department of Theoretical Physics and Gravitational Wave Science Center,  \\
24 quai E. Ansermet, CH-1211 Geneva 4, Switzerland}
\affiliation[b]{Dipartimento di Fisica, Università degli Studi di Torino, via P. Giuria, 1 10125 Torino, Italy}

\emailAdd{andrea.ianniccari@unige.ch}

\emailAdd{lorenzo.lobianco@edu.unito.it}

\emailAdd{antonio.riotto@unige.ch}

\abstract{Black holes regain   their static configuration by emitting ringdown gravitational waves, whose amplitude 
decays in  time following a power law at fixed spatial positions. We show that the nonlinear decay power law may be obtained by  simple scattering calculations using the in-in formalism and argue that the nonperturbative law should be $t^{-2\ell-1}$, where $\ell$ is the multipole of the propagating spherical gravitational wave.   }

\begin{document}
\maketitle
\flushbottom
\section{Introduction}
The advent of gravitational wave (GW) astronomy has ushered in a new era for probing the strong-field regime of gravity with unprecedented precision \cite{Berti:2015itd,Berti:2018vdi,Franciolini:2018uyq,LIGOScientific:2020tif}. Current ground-based detectors—such as LIGO, VIRGO, and KAGRA—and the upcoming space-based observatory LISA are rapidly achieving the sensitivity required to examine the post-merger, or ringdown, phase of black hole (BH) collisions in detail \cite{Berti:2005ys,KAGRA:2013rdx,LIGOScientific:2016aoc,KAGRA:2021vkt,LIGOScientific:2023lpe}. A central aspect of this effort is to analyze the behavior of BH perturbations at late times, when both linear and nonlinear effects imprint distinctive signatures on the outgoing gravitational radiation.

Traditionally, our understanding of BH ringdowns has been rooted in linear perturbation theory. According to this approach, BHs settle into equilibrium by emitting characteristic damped oscillations—known as quasinormal modes (QNMs)—which decay exponentially, followed at late times by a slower, power-law falloff (for a review, see Ref. \cite{Berti:2025hly}). A classic example is Price's result: for massless fields on a static, spherically symmetric BH background, the late-time signal at a fixed radius diminishes as $\sim t^{-2\ell-3}$, where $\ell$ is the multipole order \cite{Price:1971fb,Price:1972pw,Leaver:1986gd,Gundlach:1993tp,Gundlach:1993tn,Ching:1995tj,Chandrasekhar:1975zza,Martel:2005ir,Barack:1998bw,Berti:2009kk}.

Considerable progress has been made in refining theoretical predictions for both QNM spectra and late-time decay tails \cite{Ching:1994bd,Krivan:1996da,Krivan:1997hc,Krivan:1999wh,Burko:2002bt,Burko:2007ju,Hod:2009my,Burko:2010zj,Racz:2011qu,Zenginoglu:2012us,Burko:2013bra,Baibhav:2023clw,Rosato:2025rtr}. These linear predictions have been corroborated by numerical relativity simulations and increasingly precise GW observations.

In recent years, research attention has shifted towards the nonlinear domain of BH perturbations \cite{Gleiser:1995gx,Gleiser:1998rw,Campanelli:1998jv,Garat:1999vr,Zlochower:2003yh,Brizuela:2006ne,Brizuela:2007zza,Nakano:2007cj,Brizuela:2009qd,Ripley:2020xby,Loutrel:2020wbw,Pazos:2010xf,Sberna:2021eui,Redondo-Yuste:2023seq,Mitman:2022qdl,Cheung:2022rbm,Ioka:2007ak,Kehagias:2023ctr,Khera:2023oyf,Bucciotti:2023ets,Spiers:2023cip,Ma:2024qcv,Zhu:2024rej,Redondo-Yuste:2023ipg,Bourg:2024jme,Lagos:2024ekd,Perrone:2023jzq,Kehagias:2024sgh,Bucciotti:2025rxa,bourg2025quadraticquasinormalmodesnull,BenAchour:2024skv,Kehagias:2025ntm,Perrone:2025zhy}. This line of work has revealed phenomena that transcend the linear regime, such as the emergence of second-order QNMs and their corresponding power-law decay tails. These nonlinear features are of particular interest because they naturally emerge from the QNM structure governing BH ringdowns \cite{Okuzumi:2008ej,Lagos:2022otp}, thus offering a unique window into the nonlinear dynamics of gravity. Numerical studies have recently identified distinct decay laws in the Weyl scalar linked to these effects, differing notably from the classic Price tail behavior \cite{Cardoso:2024jme}. Moreover, full 3+1 numerical relativity simulations of BH mergers have reported power-law decays inconsistent with linear theory predictions \cite{DeAmicis:2024eoy,Ma:2024hzq}. These insights not only deepen our understanding of nonlinear dynamics in BH perturbation theory but also highlight the importance of enhancing GW modeling to capture their potential observational signatures.

One notable finding is that nonlinear interactions can generate a second-order $\ell = 4$ mode—sourced by quadratic coupling of $\ell = 2$ modes—which decays at late times as $\sim t^{-9}$, as found numerically in Ref.  \cite{Cardoso:2024jme}\footnote{To avoid any further  confusion, we point out that the nonlinear law quoted in Ref. \cite{Cardoso:2024jme}, $\sim t^{-10}$,  refers to  the RW gauge. One power less  is obtained for the helicity two degrees of freedom in the TT gauge \cite{Kehagias:2025xzm}.}. This points towards a more general decay pattern of the form $\sim t^{-2\ell-1}$, attributable to the evolution of quadratic QNMs in asymptotically flat spacetimes \cite{Ling:2025wfv,Kehagias:2025xzm,Kehagias:2025tqi}.

In this work, we reconsider the nonlinear tails of the QNMs from a different perspective. Using the in-in formalism, we   show: {\it i)} that the nonlinear QNM tails  can be simply understood as the result of the scattering of the  QNMs off the Newtonian potential in flat spacetime and {\it ii)} that at any order in perturbation theory the nonlinear tail decays as $\sim t^{-2\ell-1}$, reinforcing the argument that the nonlinearities in gravity are relevant and even potentially more important than the linear effects.

The paper is organized as follows. In section 2 we briefly remind the reader of the in-in formalism. In section 3, as a warm-up, we rederive Price's law for the linear tail. In section 4 we address the nonlinear tail power-law and conclude in section 5. The article is supplemented with two appendices.

\section{The in-in formalism}
\begin{figure}[t!]
\centering
  \includegraphics[width=0.25\textwidth]{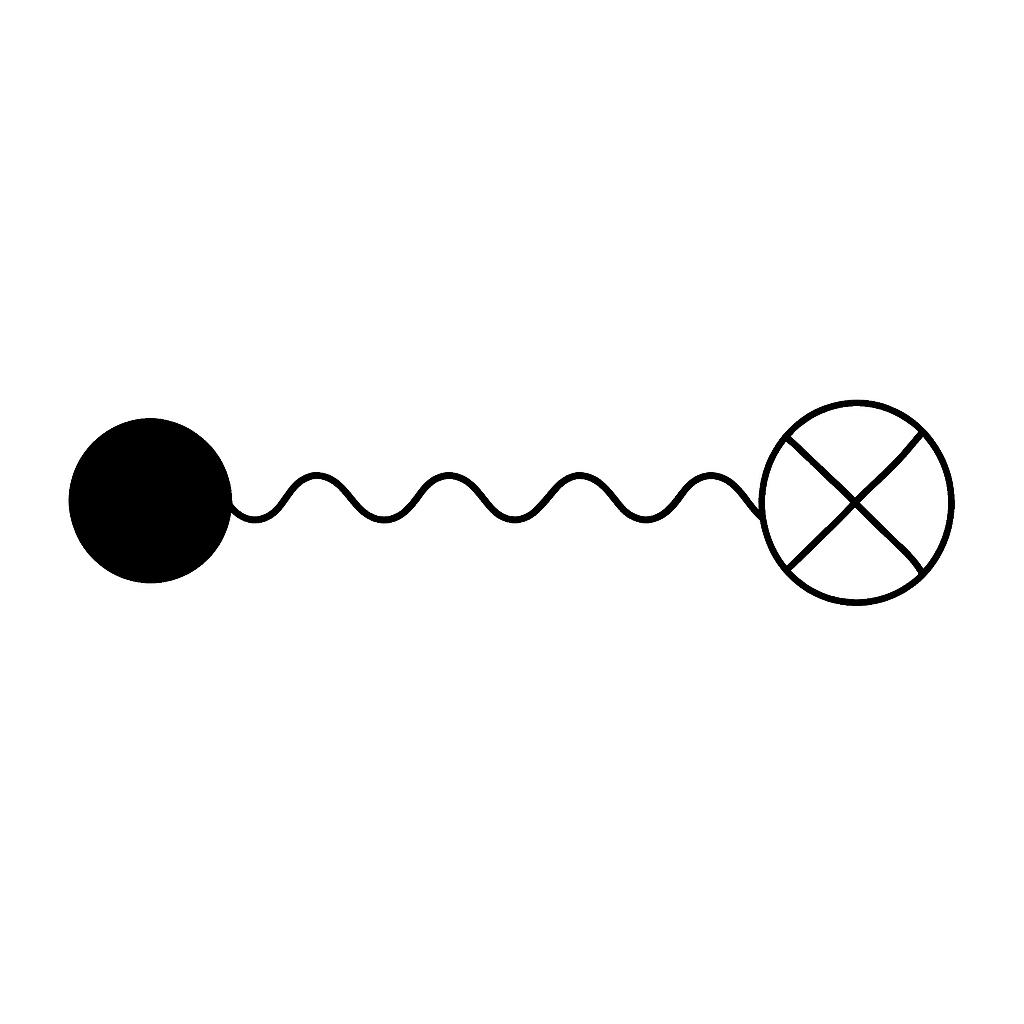}
  \caption{Tree-level graviton coupling directly to the matter source.}
  \label{fig:blob1}
\end{figure}

\begin{figure}[t!]
\centering
  \includegraphics[width=0.30\textwidth]{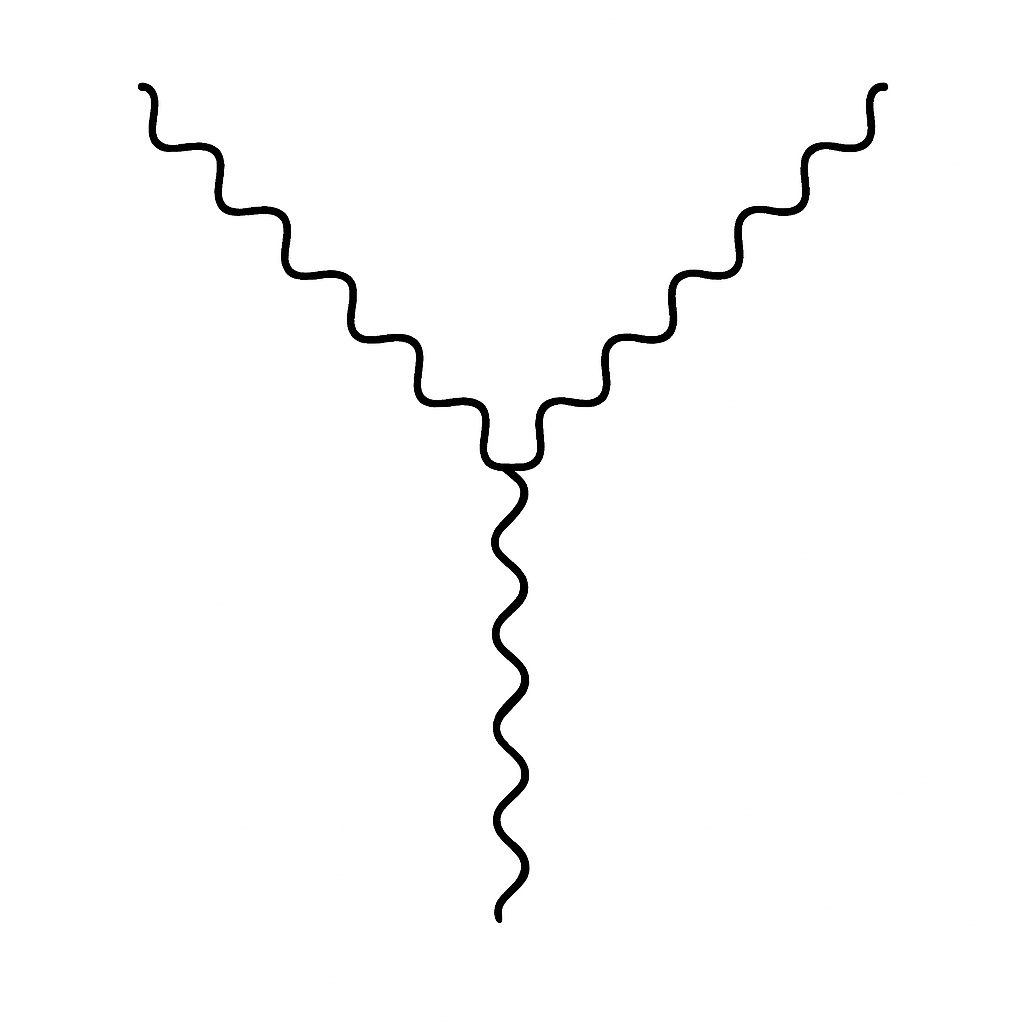}
  \caption{Three-graviton vertex.}
  \label{fig:cubictree}
\end{figure}

Our goal is to calculate the expectation value, at a given time, of the gravitational wave. Obviously, the  in-out formalism, commonly adopted in quantum field theory to calculate cross-sections from one asymptotic state to another, is not applicable: the initial and the final state are the same.  
The most appropriate extension of the field theory to deal such 
 issue it to generalize the time contour of integration to a closed-time path.
More precisely, the time integration contour is deformed to run from a given time  $t_0$ to $t$ and
back to $t_0$ \cite{Chou:1984es}. For this reason, the formalism is also dubbed in-in formalism.
Given an operator ${\cal O}$, its expectation value reads
\begin{eqnarray}\label{O}
    \langle {\cal O}(t)\rangle&=&\Big\langle\Big(Te^{-i\int^t_{t_0}dt'H_{\rm int}(t')}\Big)^{\dagger}{\cal O}(t)\Big(Te^{-i\int^t_{t_0}dt''H_{\rm int}(t'')}\Big)\Big\rangle=\nonumber\\
  &=&\sum_{N=0}^{\infty}i^N\int_{t_0}^tdt_N\int_{t_0}^{t_{N}}dt_{N-1}...\int_{t_0}^{t_2}dt_1\nonumber\\
    &\cdot&\Big\langle\Big[H_{\rm int}(t_1), \Big[H_{\rm int}(t_2),...\Big[H_{\rm int}(t_N),{\cal O}(t)\Big]...\Big]\Big]\Big\rangle,
\end{eqnarray}
where $T$ is the time-ordering operator.
Each term in the series can be identified, as we shall see, with a given Feynman diagram. In particular, starting from the  Schwarzschild metric of a spinless BH with mass $M$

\begin{equation}
    ds^2=-\left(1-\frac{2M}{r}\right)dt^2 +\left(1-\frac{2M}{r}\right)^{-1}dr^2 +r^2d\Omega^2,
\end{equation}
as long as we are far enough from the location of the BH, we can expand it for $r\gg M$. At first order in $M/r$, or at the tree-level, we get 
the classical  metric  in the Newtonian gauge
\begin{equation}
    h_{\,\mu\nu}^{\rm cl}=\begin{pmatrix}
        2M/r&0&0&0\\
        0&2M/r&0&0\\
        0&0&0&0\\
        0&0&0&0
    \end{pmatrix}
\end{equation}
and the corresponding  
diagram in Fig. \ref{fig:blob1}, where the black blob indicates the mass $M$  sourcing a classical  gravitational field. To study the nonlinear tails, we will also need the trilinear graviton vertex, for which we do not write the full matrix element  here; we will do it in the following. The corresponding Feynman diagram is given in Fig. \ref{fig:cubictree}. 

\section{The linear tail}
After an initial burst of radiation, the ringdown dominated by QNMs, the perturbation decays as a power-law tail rather than exponentially. This is the well-known Price's law \cite{Price:1972pw}. The interpretation in terms of backscattering is known and is the following. The Schwarzschild geometry has a curved spacetime that modifies how waves propagate  the spacetime curvature acts as an effective potential barrier. As a consequence, some part of the outgoing wave is backscattered off this potential. At late times, the dominant signal is not  from waves trapped near the BH, the QNMs,  but rather from backscattered waves far from the BH, scattered off the asymptotically decaying potential.
This is why the tail decays as a power law.

The goal of this section is to formalize what described above in terms of Feynman diagrams, see Fig. \ref{fig:blob1}. In this sense, it is preparatory to the next section where we will deal with the nonlinear tails.
Our starting point is the definition of the graviton plane wave in Minkowski  spacetime
\begin{equation}\label{gravv}
    h_{\mu\nu}(t,\vec{x})=\int \frac{d^3k}{(2\pi)^{3/2}}\sum_{s}\left[\frac{e^{i\vec{k}\cdot\vec{x}-i\omega t}}{\sqrt{2\omega}}\epsilon^s_{\mu\nu}a^s(\vec{k})+{\rm h.c.}\right],
\end{equation}
where $s=($+,$\times$) indicate the two graviton helicities with polarization vector $\epsilon^s_{\mu\nu}$.
Using the spherical harmonic expansion of the plane waves
\begin{equation}
    e^{i\vec{k}\cdot\vec{x}}=4\pi\sum_{\ell m}i^{\ell}j_\ell(kr)Y^*_{\ell m}(\hat{r})Y_{\ell m}(\hat{k}),
\end{equation}
we can write
\begin{equation}
     h_{\mu\nu}(t,\vec{x})=4\pi \int \frac{dk\;k^2}{(2\pi)^{3/2}\sqrt{2\omega}}\sum_s\sum_{\ell m}\Big[i^{\ell}j_\ell(kr)Y^*_{\ell m}(\theta,\phi)a^s_{\ell m}(k)\epsilon_{\mu\nu}^se^{-i\omega t}+{\rm h.c.}\Big],
\end{equation}
where we have defined the annihilation operator
\begin{equation}
    a_{\ell m}^s(k)=\int d\Omega_kY_{\ell m}(\hat{k})a^s(\vec{k}).
\end{equation}
Since the graviton is massless we can  replace the momentum $k$ with its frequency $\omega$, so that the 
 creation and annihilation operators  obey the following commutator relation
 
\begin{equation}\label{algvva}
    \left[\hat{a}^{s}_{\ell m}(\omega),\hat{a}^{\dagger s'}_{\ell'm'}(\omega')\right]=(2\pi)^3\frac{\delta(\omega-\omega')}{\omega^2}\delta_{\ell \ell'}\delta_{mm'}\delta^{ss'},
\end{equation}
where the $\omega^2$ at the denominator will cancel the measure in the three-dimensional Fourier transform of the quantized graviton.

The next step is to consider the 
quadratic  Pauli-Fierz action for the kinetic part of the graviton
\begin{eqnarray}\label{kinS}
    S &=& \frac{1}{16 \pi G} \int d^4 x \Bigg[ 
-\frac{1}{4} \partial_\rho h_{\mu \nu}\partial^\rho h^{{\rm cl}\,\mu \nu}
+ \frac{1}{2} \partial_\rho h_{\mu \nu}\partial^\nu h^{{\rm cl}\,\rho \mu}\nonumber\\
&+& \frac{1}{4} \partial_\mu h\partial^\mu h_{\rm cl}
- \frac{1}{2} \partial_\nu h^{\mu \nu}\partial_\mu h_{\rm cl}
+ {\rm perm.}\Bigg],
\end{eqnarray}
where we treat $h_{\mu\nu}$ as a quantized graviton and $h=h^\mu_\mu$. Here $h^{\rm cl }_{\mu\nu}$ is the  classical field which incorporates the information that we are not in a completely flat region of spacetime, but  in the presence of a BH.  The linear tail will be produced by the   back reaction between the gravitational wave and the curved background.

 To find the Hamiltonian, we derive the  conjugate momentum from the quadratic action
\begin{eqnarray}
    \pi_{i,\,\alpha\beta}&=&-\frac{1}{4}\partial^0h_{j,\,\alpha\beta}+\frac{1}{2}\partial_{\beta}h_{j,\,\alpha}^0+\frac{1}{4}\partial^0h_j\eta_{\alpha\beta}-\frac{1}{2}\partial_{\beta}h_j\eta^0_{\alpha}-\frac{1}{4}\partial^0h_{j,\,\alpha\beta}\nonumber\\
    &+&\frac{1}{2}\partial_{\alpha}h_{j,\,\beta}^0+\frac{1}{4}\partial^0h_j\eta_{\alpha\beta}-\frac{1}{2}\partial_{\nu}h_j^{0\nu}\eta_{\alpha\beta},
\end{eqnarray}
where $i\neq j$ can represent the graviton or, alternatively,  its classical value.
The corresponding Hamiltonian density reads
\begin{equation}
\begin{split}
    \mathcal{H}_{\rm kin}(t,\vec{x})=\pi_{\alpha\beta}\partial_0 h^{\alpha\beta}+\pi_{{\rm cl}\, \alpha\beta}\partial_0 h^{{\rm cl}\,\alpha\beta}-\mathcal{L}_{\rm kin}(t,\vec{x})
\end{split}
\end{equation}
so that the  final Hamiltonian for the quadratic vertex is the following
\begin{eqnarray}
    H_{\rm kin}(t)&=&4\pi\sqrt{G}\int  \,dr\,r^2d\Omega   \;
    \Bigg[\pi_{\alpha\beta}\partial_0 h^{\alpha\beta}+\pi_{{\rm cl}\, \alpha\beta}\partial_0 h^{{\rm cl}\,\alpha\beta}- \Bigg[-\frac{1}{4} \partial_\rho h_{\mu \nu}\partial^\rho h^{{\rm cl}\, \mu \nu}
\nonumber\\
&+& \frac{1}{2} \partial_\rho h_{\mu \nu}\partial^\nu h^{{\rm cl}\,\rho \mu}
+ \frac{1}{4} \partial_\mu h\partial^\mu h_{{\rm cl}}
- \frac{1}{2} \partial_\nu h^{\mu \nu}\partial_\mu h_{{\rm cl}}-\frac{1}{4} \partial_\rho h_{{\rm cl}\,\mu \nu}\partial^\rho h^{\mu \nu}\nonumber\\
&+& \frac{1}{2} \partial_\rho h_{{\rm cl}\,\mu \nu}\partial^\nu h^{\rho \mu}
+ \frac{1}{4} \partial_\mu h_{{\rm cl}}\partial^\mu h
- \frac{1}{2} \partial_\nu h_{{\rm cl}}^{\,\mu \nu}\partial_\mu h\Bigg]\Bigg].
\end{eqnarray}
We  take for simplicity a graviton that propagates along the $x$-axis in cartesian coordinates, with momentum
\begin{equation}
    k^{\mu}= \left(\omega,\omega,0,0\right).
\end{equation}
The corresponding polarization tensors in TT gauge are
\begin{equation}
    \epsilon^+_{\mu\nu}=\begin{pmatrix}
        0 & 0 & 0 & 0\\
        0 & 0 & 0 & 0\\
        0 & 0 & 1 & 0\\
        0 & 0 & 0 & -1
    \end{pmatrix}
\end{equation}
and 
\begin{equation}
    \epsilon^\times_{\mu\nu}=\begin{pmatrix}
        0 & 0 & 0 & 0\\
        0 & 0 & 0 & 0\\
        0 & 0 & 0 & 1\\
        0 & 0 & 1 & 0
    \end{pmatrix}.
\end{equation}
In spherical polar coordinates, which we will use in the following, the  four-momentum vector becomes
\begin{equation}
    k^{\mu}= \left(\omega,\omega \cos\phi\sin\theta,\frac{\omega\cos\theta\cos\phi}{r},\frac{-\omega\csc\theta\sin\phi}{r}\right)
\end{equation}
so that 
\begin{equation}
    \epsilon^+_{\mu\nu}=\begin{pmatrix}
        0 & 0 & 0 & 0\\
        0&(\sin\theta \sin\phi)^2-\cos^2\theta&r \sin\theta \sin^2\phi \cos\theta+r\cos\theta \sin\theta&r\sin^2\theta \sin\phi \cos\phi\\0&r \sin\theta \sin^2\phi \cos\theta+r\cos\theta \sin\theta&r^2\sin^2\phi \cos^2\theta -r^2 \sin^2\theta &r^2\sin\theta \cos\theta \sin\phi \cos\phi\\
        0&r\sin^2\theta \sin \phi \cos\phi&r^2\sin\theta \cos\theta \sin\phi \cos\phi&r^2\sin^2\theta \cos^2\phi
    \end{pmatrix}
\end{equation}
and
\begin{equation}
    \epsilon^\times_{\mu\nu}=\begin{pmatrix}
        0 & 0 & 0 & 0\\
       0 &2\sin\theta \sin\phi \cos\theta& -r \sin\phi (\sin^2\theta-\cos^2\theta)&r \cos\theta \sin\theta \cos\phi \\
      0& -r \sin\phi (\sin^2\theta-\cos^2\theta)& -2r^2 \cos\theta \sin\theta \sin\phi&-r^2 \sin^2\theta \cos\phi\\
       0&r \cos\theta \sin\theta \cos\phi&-r^2 \sin^2\theta \cos\phi&0
    \end{pmatrix}.
\end{equation}
The corresponding  kinetic Hamiltonian reads
\begin{eqnarray}
H_{\rm kin}(t)&=&4\pi\sqrt{G}\int\,dr\,r^2d\Omega\;  \;\Bigg[\int\frac{d\omega\;\omega^2}{(2\pi)^{3/2}\sqrt{2\omega}}\sum_{\ell,m}i^{\ell}j_\ell(\omega r)Y_{\ell m}^*(\theta,\phi)e^{-i\omega t }\nonumber \\
&\cdot&\left(a^+(\omega)\frac{M\omega}{r^2}f(\theta,\phi)+a^\times(\omega)\frac{M\omega}{r^2}g(\theta,\phi)\right)+{\rm h.c.}\Bigg],
\label{Hkin}
\end{eqnarray}
where
\begin{eqnarray}
f(\theta,\phi)&=&\frac{3  \Big[ 3 \cos3\phi \sin\theta + \cos\phi [ \sin\theta - 2 \left( -3 + \cos 2\phi \right) \sin3\theta ] \Big]}{16},\nonumber \\
g(\theta,\phi)&=&- 6  \cos\theta \cos\phi \sin^2\theta \sin\phi.
\end{eqnarray}
Substituting  Eq.  (\ref{Hkin}) in Eq.  (\ref{O}), we get (see Fig. \ref{fig:blob1})
\begin{eqnarray}\label{vevh}
\langle h\rangle&\equiv&\sum_s\langle\epsilon^{s\mu\nu} h_{\mu\nu}\rangle=-i\int_{t_0}^{t} dt' \Bigg\langle\left[\frac{\sqrt{G
}\;\omega^{3}a_{\omega}^{s\dagger}}{\sqrt{2\omega}}j_\ell(\omega y)e^{i\omega t},H_{\rm kin}(t')\right]\Bigg\rangle\nonumber\\
&\simeq&-i\;G\int_{t_{0}}^{t} dt'\int^{\infty} dr \;r^2d\Omega\int_{0}^{\infty}\frac{d\omega'\;\omega'^{2}}{\sqrt{2\omega'}}\frac{\omega^3}{\sqrt{2\omega}}\frac{\delta(\omega-\omega')}{\omega^2}j_{\ell}(\omega' r)j_\ell(\omega y)Y_{\ell m}^{*}(\theta,\phi)\nonumber\\
&\cdot&e^{-i\omega't'}e^{i\omega t}\frac{M\omega'}{r^2}f(\theta,\phi).
\end{eqnarray}
From now on for simplicity, we just consider the $+$ polarization.
Expanding the spherical Bessel functions for small arguments $\omega r\ll1$ and $\omega y\ll1$, we get
\begin{eqnarray}
    \langle h \rangle&=&-iG\int_{t_0}^{t} dt'\;\int dr\;r^2d\Omega\Big[j_\ell(\omega r)j_\ell(\omega y)e^{-i\omega (t'-t)}\Big]\omega^3\frac{f(\theta ,\phi)M}{r^2}Y^*_{\ell m}(\theta,\phi)\nonumber\\
    &\simeq&-iG\int_{0}^{t_0-t} d\tau\int dr\;r^2d\Omega\;\,\omega^{2\ell+3}e^{-i\omega \tau}\frac{f(\theta,\phi)M}{r^2}(yr)^\ell Y^*_{\ell m}(\theta,\phi),
\end{eqnarray}
where we have performed a change of variable $\tau=(t'-t)$.  Performing the time integral we obtain
\begin{eqnarray}
         \langle h \rangle= G \int \;dr\;r^2 d\Omega\;\omega^{2\ell+2}e^{-i\omega \tau}\frac{f(\theta,\phi)}{r^2}M(yr)^\ell Y^*_{\ell m}(\theta,\phi).
\end{eqnarray}
Finally, we need to Fourier transform the frequency part of the integral in order to get the time behavior
\begin{equation}
    \int \;dr\,r^2 d\Omega\;\frac{M(yr)^\ell}{r^2}f(\theta,\phi)Y_{\ell m}^*(\theta,\phi)\int_0^{\infty}\frac{d\omega}{2\pi}\omega^{2\ell+2}e^{-i\omega t}.
\end{equation}
Since the function is analytical in the complex plane, we can rotate the integration path along the complex axis using the residual theorem and the Jordan lemma, to get 
\begin{equation}
    \langle h \rangle\simeq\int \;dr\,r^2 d\Omega\;\frac{M(yr)^\ell}{r^2}f(\theta,\phi)Y_{\ell m}^*(\theta,\phi)\int_0^{-i\infty}\frac{d\omega}{2\pi}\omega^{2\ell+2}e^{-i\omega t}\simeq A(\theta,\phi)\frac{GMr^{\ell+1}y^{\ell}}{t^{2\ell+3}}.
\end{equation}
This is  Price's law tail, where  $A(\theta,\phi)$ encodes the angular dependence. This calculation shows that the Price's law can be understood by a simple scattering of the QNM off the classical gravitational Newtonian potential generated by the BH mass \cite{Price:1972pw,Ching:1995tj}. 
\section{The nonlinear tail}
\noindent
The aim of this section is to derive the nonlinear tail of a gravitational wave QNM which is generated by the interaction of two
other QMNs giving rise to a third wave and several rescattering off the gravitational potential generated by the mass of the BH.  For instance, we may think of an $\ell=4$ mode generated by two $\ell=2$ modes.
Our starting point is  the cubic Lagrangian obtained 

\begin{eqnarray}
\mathcal{L}^{(3)} &=& \sqrt{G}\;
\Big[  \frac{1}{4} h^{\alpha \beta} \partial_\alpha h^{\mu \nu} \partial_\beta h_{\mu \nu}
- \frac{1}{4} h^{\alpha \beta} \partial_\alpha h \, \partial_\beta h
+ h^{\alpha \beta} \partial_\beta h \, \partial_\mu h^{\mu}{}_{\alpha}
- \frac{1}{2} h^{\mu \nu} \partial_\alpha h \, \partial^\alpha h_{\mu \nu} \nonumber\\
& +& \frac{1}{8} h \, \partial_\mu h \, \partial^\mu h
-  \, h^{\mu \nu} \partial_\alpha h_\mu{}^{\alpha} \partial_\beta h_\nu{}^{\beta}
- h^{\mu \nu} \partial_\nu  \, h_\mu{}^{\alpha} \partial_\beta h_\alpha{}^{\beta}
+ \frac{1}{2} h \, \partial_\mu h^{\mu \nu} \, \partial_\alpha h_\nu{}^{\alpha}\nonumber \\
& +& \frac{1}{2} h^{\mu \nu} \partial^\alpha h_{\mu \nu} \, \partial_\beta h_\alpha{}^{\beta}
- \frac{1}{4} h \, \partial_\alpha h \, \partial_\beta h^{\alpha \beta}
+ \frac{1}{2} h^{\mu \nu} \partial_\alpha h_{\nu \beta} \, \partial^\beta h_\mu{}^{\alpha}
+ \frac{1}{2} h^{\mu \nu} \partial_\beta h_{\nu \alpha} \, \partial^\beta h_\mu{}^{\alpha}\nonumber \\
& - &\frac{1}{4} h \, \partial_\alpha h_{\mu \nu} \, \partial^\nu h^{\mu \alpha}
- \frac{1}{8} h \, \partial_\alpha h^{\mu \nu} \, \partial^\alpha h_{\mu \nu}
\Big],
\end{eqnarray}
where we have expanded the metric far away from the BH
\begin{equation}
    g_{\mu\nu}=\eta_{\mu\nu}+\sqrt{G}\,h_{\mu\nu}+\frac{G}{2}h^{2}_{\mu\nu}+\cdots.
\end{equation}
Using the TT gauge, the action reduces to 
\begin{eqnarray}
\label{5.2}
S_3[h] &= &4\pi\sqrt{G}\;\int dt \;dr\;r^2\;d\Omega \;
\Big[ \frac{1}{4} h^{\alpha \beta} \partial_\alpha h^{\mu \nu} \partial_\beta h_{\mu \nu} -  \, h^{\mu \nu} \partial_\alpha h_\mu{}^{\alpha} \partial_\beta h_\nu{}^{\beta}
- h^{\mu \nu} \partial_\nu  \, h_\mu{}^{\alpha} \partial_\beta h_\alpha{}^{\beta}\nonumber \\
& +& \frac{1}{2} h^{\mu \nu} \partial^\alpha h_{\mu \nu} \, \partial_\beta h_\alpha{}^{\beta}
+ \frac{1}{2} h^{\mu \nu} \partial_\alpha h_{\nu \beta} \, \partial^\beta h_\mu{}^{\alpha}
+ \frac{1}{2} h^{\mu \nu} \partial_\beta h_{\nu \alpha} \, \partial^\beta h_\mu{}^{\alpha}
\Big].
\end{eqnarray}
The integral in Eq. (\ref{5.2}) becomes
\begin{eqnarray}
    S_3[h]&=&\frac{\sqrt{G}}{4}\int dt\;dr\;r^2\;d\Omega\sum_{s_i,\ell_i,m_i}\int \frac{d\omega_1}{\sqrt{2\omega_1}}\frac{d\omega_2}{\sqrt{2\omega_2}}\frac{d\omega_3}{\sqrt{2\omega_3}}\omega_1^2\omega_2^2\omega_3^2j_{\ell_1}(\omega_1r)j_{\ell_2}(\omega_2r)j_{\ell_3}(\omega_3r)\nonumber\\
    &\cdot& Y_{\ell_1m_1}(\theta,\phi)e^{-i\omega_1t}Y^*_{\ell_2m_2}(\theta,\phi)e^{i\omega_2t}Y_{\ell_3m_3}(\theta,\phi)e^{-i\omega_3t}\cdot\Big[ \frac{1}{4} \epsilon_1^{\alpha \beta} k_{2,\,\alpha} \epsilon_{2}^{\mu \nu} k_{3,\,\beta} \epsilon_{3, \,\mu \nu}\nonumber
    \\
    &-& \epsilon_{1}^{\mu \nu} k_{2,\,\alpha} \epsilon_{2,\,\mu}^{\alpha} k_{3,\,\beta} \epsilon_{3,\,\nu}^{\beta}
- \epsilon_1^{\mu \nu} k_{2,\,\nu} \epsilon_{2,\,\mu}^{\alpha} k_{3,\,\beta} \epsilon_{3,\,\alpha}^{\beta}
+ \frac{1}{2} \epsilon_1^{\mu \nu} k_2^{\,\alpha} \epsilon_{2,\,\mu \nu} k_{3,\,\beta} \epsilon_{3,\,\alpha}^{\beta}\nonumber\\
&+&\frac{1}{2} \epsilon_1^{\mu \nu} k_{2,\,\alpha} \epsilon_{2,\,\nu \beta} k_3^{\,\beta} \epsilon_{3,\,\mu}^{\alpha}
+ \frac{1}{2} \epsilon_1^{\mu \nu} k_{2,\,\beta} \epsilon_{2,\,\nu \alpha} k_3^{\,\beta} \epsilon_{3,\,\mu}^{\alpha}+{\rm perm}(1,2,3)\Big]a(\omega_1)a^{\dagger}(\omega_2)a(\omega_3)+\cdots.\nonumber\\
&&
\label{vertex}
\end{eqnarray}
Incidentally, we notice that the conjugate momenta from the interaction $\pi_{\alpha\beta}=\delta\mathcal{L}^{(3)}/\delta(\partial_0h^{\alpha\beta})$ are vanishing in the  TT gauge. 
From now on we also consider the case of two identical QNMs, such that $\omega_1= \omega_2$.
The interaction Hamiltonian becomes 
\begin{eqnarray}
    H_{\,\,{\rm int}}^{{\rm q}\,{\rm q}\,{\rm q}}(t)&=&\frac{\sqrt{G}}{4}\int\;dr\;r^2\;d\Omega\sum_{s_i,\ell_i,m_i}\int \frac{d\omega_1}{\sqrt{2\omega_1}}\frac{d\omega_2}{\sqrt{2\omega_2}}\frac{d\omega_3}{\sqrt{2\omega_3}}\omega_1^2\omega_2^2\omega_3^2j_{\ell_1}(\omega_1r)j_{\ell_2}(\omega_2r)j_{\ell_3}(\omega_3r)\nonumber\\
     &\cdot& Y^*_{\ell_1m_1}(\theta,\phi)e^{-i\omega_1t}Y_{\ell_2m_2}(\theta,\phi)e^{i\omega_2t}Y^*_{\ell_3m_3}(\theta,\phi)e^{-i\omega_3t}\;\mathcal{L}_{\rm tens}\;\delta(\omega_1-\omega_2)a(\omega_1)a^{\dagger}(\omega_2)a(\omega_3)+\cdots,\nonumber\\
       \mathcal{L}_{\rm tens}&=&\sqrt{G}\;\Big[ \frac{1}{4} \epsilon_1^{\,\alpha \beta} k_{2,\,\alpha} \epsilon_{2}^{\,\mu \nu} k_{3,\,\beta} \epsilon_{3\,\mu \nu} - \epsilon_{1}^{\,\mu \nu} k_{2,\,\alpha} \epsilon_{2,\,\mu}^{\,\alpha} k_{3,\,\beta} \epsilon_{3,\,\nu}^{\,\beta}
- \epsilon_1^{\,\mu \nu} k_{2,\,\nu} \epsilon_{2,\,\mu}^{\alpha} k_{3,\,\beta} \epsilon_{3,\,\alpha}^{\beta}\\
&+& \frac{1}{2} \epsilon_1^{\,\mu \nu} k_2^{\,\alpha} \epsilon_{2,\,\mu \nu} k_{3,\,\beta} \epsilon_{3,\,\alpha}^{\,\beta}\nonumber
+ \frac{1}{2} \epsilon_1^{\,\mu \nu} k_{2,\,\alpha} \epsilon_{2,\,\nu \beta} k_3^{\,\beta} \epsilon_{3,\,\mu}^{\,\alpha}+ \frac{1}{2} \epsilon_1^{\,\mu \nu} k_{2,\,\beta} \epsilon_{2,\,\nu \alpha} k_3^{\,\beta} \epsilon_{3,\,\mu}^{\,\alpha}+{\rm perm}(1,2,3)\Big]\nonumber\\
&=&\frac{3\omega_1^2 +6\omega_1\omega_3}{4}.
\end{eqnarray}
Here $[\rm q\,q\,q]$ in the interaction Hamiltonian means that we are inserting three quantized gravitons.
The last passage is derived in Appendix \ref{AppendixA}. Notice that the cubic interaction introduces the Clebsch-Gordon coefficients in the action 
\begin{equation}
    \int d\Omega\; Y_{\ell_1m_1}Y_{\ell_2m_2}Y_{\ell_3m_3}= \mathcal{C}_{\ell_1\ell_2\ell_3,m_1m_2m_3},
\end{equation}
which impose  angular momentum conservation selecting  
  $m_1+m_2=m_3$ and $|\ell_1-\ell_2|\leq\ell_3\leq \ell_1+\ell_2$. Specifically, in our case, we have 

\begin{equation}
    \int d\Omega\; Y^*_{\ell_1m_1}Y_{\ell_2m_2}Y^*_{\ell_3m_3}= (-1)^{m_1+m_3}\mathcal{C}_{\ell_1\ell_2\ell_3,-m_1m_2-m_3}.
\end{equation}
\subsection{No external sources: the tadpole}
In the case of no external sources we start from 
\begin{eqnarray}
    H_{\,\,{\rm int}}^{{\rm q}\,{\rm q}\,{\rm q}}(t)&\simeq&\frac{\sqrt{G}}{4}\int^{\infty}dr\; r^2 \int_0^{\infty} \frac{d\omega_1\;\omega_1^4}{2\omega_1}\int_0^{\infty}\frac{d\omega_3}{\sqrt{2\omega_3}}\omega_3^2
    e^{-i\omega_3t}\;\mathcal{L}_{\rm tens}\;
    j_{\ell_1}(\omega_1r)j_{\ell_2}(\omega_1r)j_{\ell_3}(\omega_3r)\nonumber\\
    &\cdot& a(\omega_1)a^{\dagger}(\omega_1)a(\omega_3)+\cdots
\end{eqnarray}
and  can easily calculate the commutator
\begin{figure}[t!]
\centering
  \includegraphics[width=0.25\textwidth]{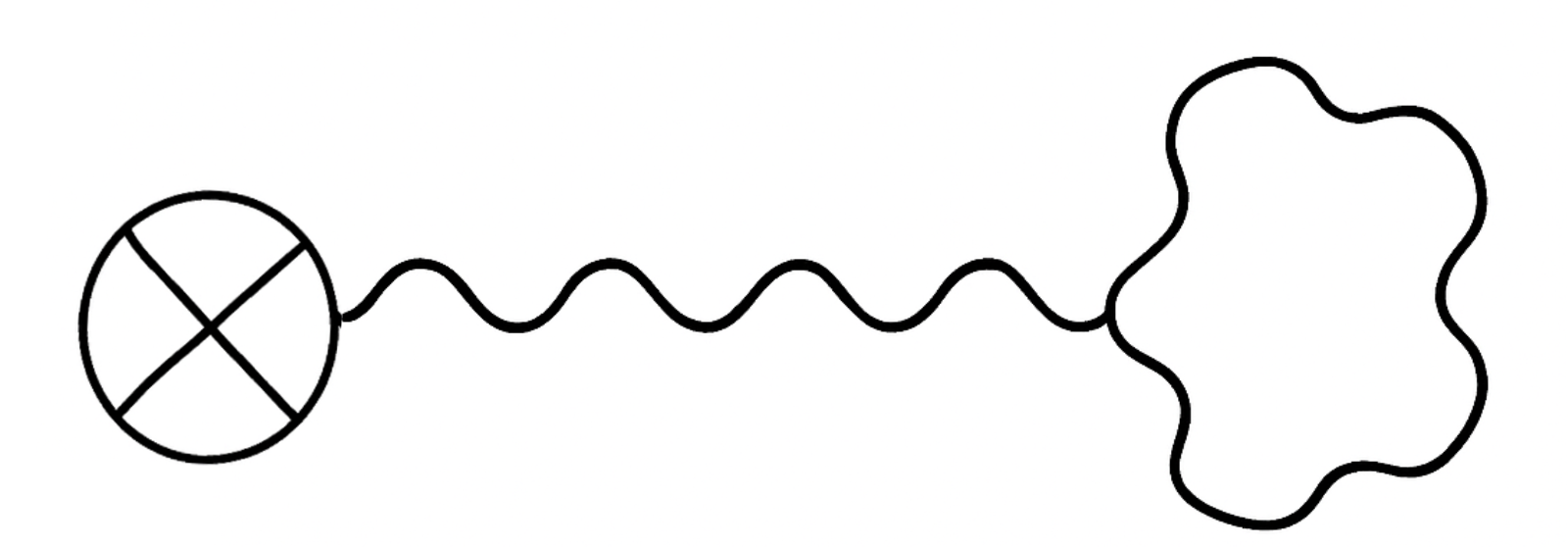}
  \caption{Tadpole diagram.}
  \label{fig:tadpole}
\end{figure}
\begin{eqnarray}
\label{Acommutator}
    A&=&\left[H_{\,\,{\rm int}}^{{\rm q}\,{\rm q}\,{\rm q}}(t'),h(t)\right]=\frac{G}{4}\int^{\infty}dr\; r^2 \int_0^{\infty} \frac{d\omega_1\;\omega_1^3}{2}\int_0^{\infty}\frac{d\omega_3}{\sqrt{2\omega_3}}\omega_3^2
    e^{-i\omega_3t'}e^{i\omega' t}\;\mathcal{L}_{\rm tens}\nonumber\\
    &\cdot& j_{\ell_1}(\omega_1r) j_{\ell_2}(\omega_1r)j_{\ell_3}(\omega_3r)j_{\ell_3}(\omega'y)
 \frac{\omega'^3}{\sqrt{\omega'}}\frac{\delta(\omega'-\omega_3)}{\omega_3^2}a_{\omega_1}a_{\omega_1}^{\dagger}\nonumber\\
    &\simeq&\int^{\infty}dr\; r^2 \; \;\int_{0}^{\infty}d\omega_1 \omega_1^5\omega'^{2}e^{-i\omega'(t'-t)}\;j_{\ell_1}(\omega_1r)j_{\ell_2}(\omega_1r)j_{\ell_3}(\omega'r)j_{\ell_3}(\omega'y)a_{\omega_1}a_{\omega_1}^{\dagger},
\end{eqnarray}
such that, at first order in the Hamiltonian
\begin{equation}\label{hhtwocrossess}
    \langle h \rangle= \int_{t_0}^{t}dt'\Big\langle\left[H_{\,\,{\rm int}}^{{\rm q}\,{\rm q}\,{\rm q}}(t'),h(t)\right]\Big\rangle.
\end{equation}
Performing the integral, we obtain
\begin{eqnarray}
   \langle h \rangle&\simeq&\frac{G}{4}\int_{t_0}^{t}dt'\int^{\infty}dr\;r^2\int_0^{\infty} \frac{d\omega_1\;\omega_1^5}{2}\omega'^2
    e^{-i\omega'(t'-t)}\;j_{\ell_1}(\omega_1r)j_{\ell_2}(\omega_1r)j_{\ell_3}(\omega'r) j_{\ell_3}(\omega' y) \nonumber\\
    &\cdot&\Big\langle a(\omega_1)a^{\dagger}(\omega_1)\Big\rangle.
\end{eqnarray}
This result corresponds diagrammatically to a tadpole, see Fig. \ref{fig:tadpole}, which  is renormalized  to zero by an appropriate counterterm. As expected, there is therefore no contribution to the nonlinear tail in pure flat space time.
\subsection{Two external sources}
\begin{figure}[t!]
\centering
  \includegraphics[width=0.25\textwidth]{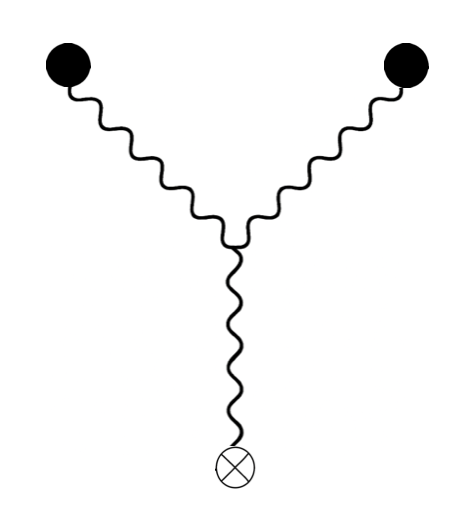}
  \caption{Two external sources.}
  \label{fig:twosources}
\end{figure}

In this subsection we provide the calculations for the diagram at order $(GM)^2$, see Fig. \ref{fig:twosources}, and we take $\ell_1=\ell_2=1$ and $\ell_3=2$ (this choice will be clear in the following), for which

\begin{equation}
    \langle h \rangle= \int_{t_0}^{t}dt_3\int_{t_0}^{t_3}dt_2\int_{t_0}^{t_2}dt_1\Big<\left[H_{\rm kin}(t_1),\left[H_{\rm kin}(t_2),\left[H_{\,\,{\rm int}}^{{\rm q}\,{\rm q}\,{\rm q}}(t_3),h(t)\right]\right]\right]\Big>.
\label{hh}
\end{equation}
The first inner commutator has been calculated already, see  Eq. (\ref{Acommutator}). 
The second commutator in Eq. (\ref{hh})
will be 
\begin{eqnarray}
      B&=&\left[ H_{\rm kin}(t_2),A\right]\simeq  \int^{\infty}dr_3\; r_3^2 \; \;\int_{0}^{\infty}d\omega_1 \omega_1^5\omega'^{2}e^{-i\omega'(t_3-t)}j_{1}^2(\omega_1r_3)j_{2}(\omega'r_3) j_{2}(\omega'y)\nonumber\\
    &\cdot&\int dr_2\; r_2^2 \; d\Omega_2\int_{0}^{\infty}\frac{d\omega_4\; \omega_4^2}{\sqrt{2\omega_4}}e^{-i\omega_4 t_2}\frac{M\omega_4}{r_2^2}\sum_{\ell_4,m_4}i^{\ell_4}Y_{\ell_4m_4}^*(\theta_2,\phi_2)j_{\ell_4}(\omega_4r_2)f(\theta_2,\phi_2)\nonumber\\
&\cdot&\left[a_{\omega_4},a_{\omega_1}a_{\omega_1}^{\dagger}\right].
\end{eqnarray}
Since
\begin{equation}
    \left[a_{\omega_4},a_{\omega_1}a_{\omega_1}^{\dagger}\right]= \frac{a_{\omega_1}\delta(\omega_4-\omega_1)}{\omega_1^2}\delta_{\ell_41}
\end{equation}
we get 
\begin{eqnarray}
      B &=&\left[ H_{\rm kin}(t_2),A\right]\simeq  \int^{\infty}dr_3\; r_3^2 \; \;\int_{0}^{\infty}d\omega_1 \omega_1^5\omega'^{2}e^{-i\omega'(t_3-t)}j_{1}^2(\omega_1r_3)j_{2}(\omega'r_3) j_{2}(\omega'y)\nonumber\\
    &\cdot&\int^{\infty}dr_2\; r_2^2 \; \frac{\; \omega_1^2}{\sqrt{2\omega_1}}j_{2}(\omega_1r_2)e^{-i\omega_1 t_2}\frac{M\omega_1}{r_2^2}\frac{a_{\omega_1}}{\omega_1^2}. 
\end{eqnarray}
Consequently, the final commutator will be
\begin{eqnarray}
      C&=&\left[ H_{\rm kin}(t_1),B\right]\simeq \int^{\infty}dr_3\; r_3^2 \; \;\int_{0}^{\infty}d\omega_1 \omega_1^5\omega'^{2}e^{-i\omega'(t_3-t)}j_{1}^2(\omega_1r_3)j_{2}(\omega'r_3) j_2(\omega'y)\nonumber\\
    &\cdot&\int^{\infty}dr_2\; r_2^2 \; \frac{\;j_{2}(\omega_1r_2)}{\sqrt{2\omega_1}}e^{-i\omega_1 t_2}\frac{M\omega_1}{r_2^2}\int dr_1\; r_1^2 \; d\Omega_1\nonumber\\
&\cdot&\int_{0}^{\infty}\frac{d\omega_5\; \omega_5^2}{\sqrt{2 \omega_5}}e^{i\omega_5t_1}\frac{M\omega_5}{r_1^2}\sum_{\ell_5,m_5}i^{\ell_5}Y_{\ell_5m_5}(\theta_1,\phi_1)j_{\ell_5}(\omega_5 r_1)f(\theta_1,\phi_1)\left[a_{\omega_1},a_{\omega_5}^{\dagger}\right].
\end{eqnarray}
Using 
\begin{equation}
    \left[a_{\omega_1},a_{\omega_5}^{\dagger}\right]= \frac{\delta(\omega_1-\omega_5)}{\omega_1^2}\delta_{\ell_51},
\end{equation}
we get 
\begin{eqnarray}
       C&\simeq & \int^{\infty}dr_3\; r_3^2 \; \int^{\infty}dr_2\; r_2^2 \;  \int^{\infty}dr_1\; r_1^2 \; \int_{0}^{\infty}d\omega_1 \omega_1^6 \omega'^{2}j_{2}(\omega'r_3)j_{2}(\omega'y)\nonumber\\
       &\cdot &j_{1}^2(\omega_1r_3)j_{1}(\omega_1r_2)j_{1}(\omega_1r_1)\frac{M^2}{r_1^2r_2^2}e^{-i\omega'(t_3-t)}e^{-i\omega_{1}t_2}e^{i\omega_{1}t_1}.
\end{eqnarray}
Finally we can write the vacuum expectation value of the graviton as 
\begin{eqnarray}
  \langle h \rangle &\simeq& \int_{t_0}^{t}dt_3\; \int_{t_0}^{t_3}dt_2\int_{t_0}^{t_2}dt_1 \int^{\infty}dr_3\; r_3^2 \;  \int^{\infty}dr_2\;  \;  \int^{\infty}dr_1\; \nonumber\\&\cdot&\int_{0}^{\infty}d\omega_1 \omega_1^6 j_{1}^2(\omega_1r_3)j_{1}(\omega_1r_2)j_{1}(\omega_1r_1)e^{i\omega_1 t_1}e^{-i\omega_1 t_2}  \nonumber\\&\cdot&
  \int_{0}^{-i\infty}d\omega'\omega'^{2}j_{2}(\omega'r_3)j_{2}(\omega'y)
 e^{-i\omega'(t_3-t)} M^2 \nonumber\\
 &&
\end{eqnarray}
where we Fourier transformed the frequency part of the integral in order to get the time behavior.
Performing the time integral we get 
\begin{eqnarray}\label{hvev2leg}
  \langle h \rangle &\simeq& \int_{0}^{t_0-t}d\tau_3\;\tau_3  \int^{\infty}dr_3\; r_3^2 \;  \int^{\infty}dr_2\;  \;  \int^{\infty}dr_1\; \nonumber\\&\cdot&\int_{0}^{\infty}d\omega_1 \omega_1^5 j_{1}^2(\omega_1r_3)j_{1}(\omega_1r_2)j_{1}(\omega_1r_1)  \nonumber\\&\cdot&
  \int_{0}^{-i\infty}d\omega'\omega'^{2}j_{2}(\omega'r_3)j_{2}(\omega'y)
 e^{-i\omega'\tau_3} M^2 \nonumber\\
 &&
 \label{hmed2s}
\end{eqnarray}
where  $\tau_3=(t_3-t)$.
Here we can recognize the usual Price Green function \cite{Andersson:1996cm}
\begin{equation}
    G(r_3,\tau_3,y) = \int_{0}^{-i\infty}d\omega'\omega'^{2}j_{\ell}(\omega'r_3)j_\ell(\omega'y)
 e^{-i\omega'\tau_3}.
\label{Andersonlaw}
\end{equation}
Inserting this result  in Eq. (\ref{hvev2leg}) one gets at $(GM)^2$ order
\begin{equation}
    \langle h\rangle\sim \frac{(GM)^2}{t^{5}}.
\end{equation}
We now see that there is a specific selection rule which links the value of $\ell_3\leq 2$ of the final state with respect to the perturbative order in $GM$. 
In  the static limit $t_3\gg r_3$ \cite{Kehagias:2025tqi}
\begin{equation}
     G(r_3,t_3,y)\sim \frac{P_{\ell_3}(\chi)}{t_3},
\label{Pleg}
\end{equation}
where we have defined $\rm AdS_2$ invariant distance
\begin{equation}
    \chi=\frac{-(t-t_3)^2+(r_3-y)^2}{2r_3y},
\end{equation}
so that  we can simplify the integrals (returning to the generic case $\ell_1=\ell_2$ and $\ell_3=(\ell_1+\ell_2)$) as 
\begin{eqnarray}\label{hvevPl}
  \langle h \rangle &\simeq&M^2 \int_{0}^{t_0-t}d\tau_3\;\tau_3  \int^{\infty}dr_3\; r_3^2 \;  \int^{\infty}dr_2\;  \;  \int^{\infty}dr_1\; \nonumber\\&\cdot&\int_{0}^{\infty}d\omega_1 \omega_1^5 j_{\ell_1}(\omega_1r_3)j_{\ell_2}(\omega_1r_3)j_{\ell_1}(\omega_1r_2)j_{\ell_1}(\omega_1r_1)\frac{P_{\ell_3}(\chi)}{\tau_3}. \nonumber\\
 &&
\end{eqnarray}
Far away from the BH we can confuse the positions of the three vertices, getting
\begin{eqnarray}
  \langle h \rangle&\simeq& \lim_{r_1,\;r_2,\;r_3\to\infty}M^2\int_{0}^{t_0-t}d\tau_3\;  \int^{\infty}dr_3\; r_3^2 \;\int^{\infty}dr_2\;  \int^{\infty}dr_1  \; \delta(r_1-r_2)\delta(r_2-r_3)\nonumber\\
  & \cdot& \int_{0}^{\infty}d\omega_1 \omega_1^5 j_{\ell_1}(\omega_1r_3)j_{\ell_2}(\omega_1r_3)j_{\ell_1}(\omega_1r_2)j_{\ell_1}(\omega_1r_1)P_{\ell_3}(\chi)      
\end{eqnarray}
so that 
\begin{equation}
\begin{split}
    \langle h\rangle \sim \int_{0}^{t_0-t}d\tau_3\; \int^{\infty}dr\ P_{\ell_3}(\chi)\left(\frac{GM}{r}\right)^2,
\end{split}
\end{equation}
where we have expanded the spherical Bessel function $j_{\ell_i}(\omega_1 r)$ for $\omega_1 r\gg1$. 
Now, following the Ref. \cite{Kehagias:2025tqi} it is easy to show that from the orthogonality's properties of the Legendre polynomial, the first non-zero contribution will be given by $\ell_3\leq 2$.

\subsection{Four external sources}
\begin{figure}[t!]
\centering
  \includegraphics[width=0.30\textwidth]{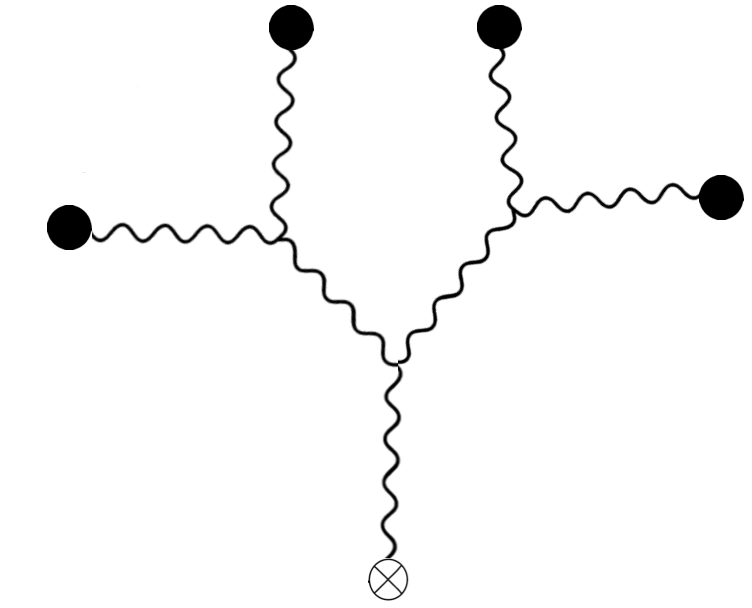}
  \caption{Four external sources.}
  \label{fig:foursources}
\end{figure}

In this section we provide the calculations for the diagram at order $(GM)^4$, see Fig. \ref{fig:foursources}. We take $\ell_1=\ell_2=2$ and $\ell_3=4$ (this choice, as did in the previous section, will be clear in the following).
We  need to calculate the following vacuum expectation value for the graviton
\begin{equation}
    \langle h \rangle= \int_{t_0}^{t}dt_3\int_{t_0}^{t_3}dt_2\int_{t_0}^{t_2}dt_1\Big<\left[H_{\,\,{\rm int}}^{{\rm q}\,{\rm cl}\,{\rm cl}}(t_1),\left[H_{\,\,{\rm int}}^{{\rm q}\,{\rm cl}\,{\rm cl}}(t_2),\left[H_{\,\,{\rm int}}^{{\rm q}\,{\rm q}\,{\rm q}}(t_3),h(t)\right]\right]\right]\Big>
\label{h}
\end{equation}
where $[\rm q \,cl\,cl]$ in the interaction Hamiltonian means that we are inserting one quantized graviton and two classical sources. 
The first inner commutator has been calculated already, and it is Eq. (\ref{Acommutator}),
\begin{equation}
    A=\left[H_{\,\,{\rm int}}^{{\rm q}\,{\rm q}\,{\rm q}}(t_3),h(t)\right]=\int^{\infty}dr_3\; r_3^2 \; \;\int_{0}^{\infty}d\omega_1 \omega_1^5\omega'^{2}e^{-i\omega'(t_3-t)}\;j_2^2(\omega_1r_3)j_4(\omega'r_3)j_4(\omega'y)a_{\omega_1}a_{\omega_1}^{\dagger}.
\end{equation}
The second commutator in Eq. (\ref{h}) reads
\begin{eqnarray}
      B&=&\left[H_{\,\,{\rm int}}^{{\rm q}\,{\rm cl}\,{\rm cl}}(t_2),A\right]=  \int^{\infty}dr_3\; r_3^2 \; \;\int_{0}^{\infty}d\omega_1 \omega_1^5\omega'^{2}e^{-i\omega'(t_3-t)}j_2^2(\omega_1r_3)j_4(\omega'r_3) j_4(\omega'y)\nonumber\\
    &\cdot&\int dr_2\; r_2^2 \; d\Omega_2\int_{0}^{\infty}\frac{d\omega_4\; \omega_4^2}{\sqrt{2\omega_4}}e^{-i\omega_4 t_2}\frac{M^2\omega_4}{r_2^3}\sum_{\ell_4,m_4}i^{\ell_4}Y_{\ell_4m_4}^*(\theta_2,\phi_2)j_{\ell_4}(\omega_4r_2)f(\theta_2,\phi_2)\nonumber\\
&\cdot&\left[a_{\omega_4},a_{\omega_1}a_{\omega_1}^{\dagger}\right].
\end{eqnarray}
The calculation of $H_{\,\,{\rm int}}^{{\rm q}\,{\rm cl}\,{\rm cl}}(t_2)$ is analyzed in Eq. (\ref{HintNLP}).
Since
\begin{equation}
    \left[a_{\omega_4},a_{\omega_1}a_{\omega_1}^{\dagger}\right]= \frac{a_{\omega_1}\delta(\omega_4-\omega_1)}{\omega_1^2}\delta_{\ell_42}
\end{equation}
we get 
\begin{eqnarray}
      B&=&\left[ H_{\,\,{\rm int}}^{{\rm q}\,{\rm cl}\,{\rm cl}}(t_2),A\right]\simeq  \int^{\infty}dr_3\; r_3^2 \; \;\int_{0}^{\infty}d\omega_1 \omega_1^5\omega'^{2}e^{-i\omega'(t_3-t)}j_2^2(\omega_1r_3)j_4(\omega'r_3) j_4(\omega'y)\nonumber\\
    &\cdot&\int^{\infty}dr_2\; r_2^2 \; \frac{\; \omega_1^2}{\sqrt{2\omega_1}}j_2(\omega_1r_2)e^{-i\omega_1 t_2}\frac{M^2\omega_1}{r_2^3}\frac{a_{\omega_1}}{\omega_1^2}. 
\end{eqnarray}
Therefore the final commutator will be
\begin{eqnarray}
      C&=&\left[ H_{\,\,{\rm int}}^{{\rm q}\,{\rm cl}\,{\rm cl}}(t_1),B\right]\simeq  \int^{\infty}dr_3\; r_3^2 \; \;\int_{0}^{\infty}d\omega_1 \omega_1^5\omega'^{2}e^{-i\omega'(t_3-t)}j_2^2(\omega_1r_3)j_4(\omega'r_3) j_4(\omega'y)\nonumber \\
    &\cdot&\int^{\infty}dr_2\; r_2^2 \; \frac{\;j_2(\omega_1r_2)}{\sqrt{2\omega_1}}e^{-i\omega_1 t_2}\frac{M^2\omega_1}{r_2^3}\int dr_1\; r_1^2 \; d\Omega_1\nonumber\\
&\cdot&\int_{0}^{\infty}\frac{d\omega_5\; \omega_5^2}{\sqrt{2 \omega_5}}e^{i\omega_5t_1}\frac{M^2\omega_5}{r_1^3}\sum_{\ell_5,m_5}i^{\ell_5}Y_{\ell_5m_5}(\theta_1,\phi_1)j_{\ell_5}(\omega_5 r_1)f(\theta_1,\phi_1)\left[a_{\omega_1},a_{\omega_5}^{\dagger}\right].
\end{eqnarray}
Since now 
\begin{equation}
    \left[a_{\omega_1},a_{\omega_5}^{\dagger}\right]= \frac{\delta(\omega_1-\omega_5)}{\omega_1^2}\delta_{\ell_52},
\end{equation}
we get 
\begin{eqnarray}
       C&\simeq & \int^{\infty}dr_3\; r_3^2 \; \int^{\infty}dr_2\; r_2^2 \;  \int^{\infty}dr_1\; r_1^2 \; \int_{0}^{\infty}d\omega_1 \omega_1^6 \omega'^{2}j_4(\omega'r_3)j_4(\omega'y)\nonumber\\
       &\cdot &j_2^2(\omega_1r_3)j_2(\omega_1r_2)j_2(\omega_1r_1)\frac{M^4}{r_1^3r_2^3}e^{-i\omega'(t_3-t)}e^{-i\omega_{1}t_2}e^{i\omega_{1}t_1}.
\end{eqnarray}
Finally, we can write the vacuum expectation value of the graviton as 
\begin{eqnarray}
  \langle h \rangle &\simeq& \int_{t_0}^{t}dt_3\; \int_{t_0}^{t_3}dt_2\int_{t_0}^{t_2}dt_1 \int^{\infty}dr_3\; r_3^2 \;  \int^{\infty}\frac{dr_2}{r_2}\;  \;  \int^{\infty}\frac{dr_1}{r_1} \; \nonumber\\&\cdot&\int_{0}^{\infty}d\omega_1 \omega_1^6 j_2^2(\omega_1r_3)j_2(\omega_1r_2)j_2(\omega_1r_1)e^{i\omega_1 t_1}e^{-i\omega_1 t_2}  
  \int_{0}^{-i\infty}d\omega'\omega'^{2}j_4(\omega'r_3)j_4(\omega'y)
 e^{-i\omega'(t_3-t)} M^4 \nonumber\\
 &&
\end{eqnarray}
where we Fourier transformed the frequency part of the integral in order to get the time behavior.
Performing the integrals over the times we get, at leading order in $t$,
\begin{eqnarray}
  \langle h \rangle&\simeq &\int_{0}^{t_0-t}d\tau_3\; \tau_3 \int^{\infty}dr_3\; r_3^2 \; \int^{\infty}\frac{dr_2}{r_2}\;  \; \int^{\infty}\frac{dr_1}{r_1} \;\int_{0}^{\infty}d\omega_1 \omega_1^5 j_2^2(\omega_1r_3)j_2(\omega_1r_2)j_2(\omega_1r_1)\nonumber\\ 
  &\cdot&\int_{0}^{-i\infty}d\omega'\omega'^{2}j_4(\omega'r_3)j_4(\omega'y)
 e^{-i\omega'\tau_3} M^4,      
\label{hvevl4}
\end{eqnarray}
where  $\tau_3=(t_3-t)$.
We can easily see that the time dependence is 
\begin{equation}
    \langle h\rangle\sim \frac{(GM)^4}{t^{9}}.
\end{equation}
As done in the previous section, we find a specific selection rule which links the value of $\ell_3\leq4$ of the final state with respect to the perturbative order in $GM$. We can plug the expression (\ref{Pleg}) in Eq. (\ref{hvevl4}) and return to the generic case $\ell_1=\ell_2$ and $\ell_3=(\ell_1+\ell_2)$ 
\begin{eqnarray}
  \langle h \rangle &\simeq&M^4 \int_{0}^{t_0-t}d\tau_3\;\tau_3  \int^{\infty}dr_3\; r_3^2 \;  \int^{\infty}\frac{dr_2}{r_2}\;  \;  \int^{\infty}\frac{dr_1}{r_1}\; \nonumber\\&\cdot&\int_{0}^{\infty}d\omega_1 \omega_1^5 j_{\ell_1}(\omega_1r_3)j_{\ell_2}(\omega_1r_3)j_{\ell_1}(\omega_1r_2)j_{\ell_1}(\omega_1r_1)\frac{P_{\ell_3}(\chi)}{\tau_3}.
\end{eqnarray}
Again, far  away from the BH we can confuse the positions of the three vertices, getting

\begin{equation}
\begin{split}
    \langle h\rangle \sim \int_{_0}^{t_0-t}d\tau_3\; \int^{\infty}dr\ P_{\ell_3}(\chi)\left(\frac{GM}{r}\right)^4,
\end{split}
\end{equation}
where we have expanded the spherical Bessel function $j_{\ell_i}(\omega_1 r)$ for $\omega_1 r\gg1$. 
As in the previous subsection, the orthogonality's properties of the Legendre polynomial, selects a  non zero result only for  $\ell_3\leq 4$. We infer the generic rule that the dominant contribution to the nonlinear tail of a QNM of multipole $\ell_3$ generated by 
the annihilation of two QNMs with multipoles $\ell_1=\ell_2$ reads

\begin{equation}
     \langle h\rangle\sim \frac{(GM)^{\ell_3}}{t^{2\ell_3+1}}.
\end{equation}
We stress  again that we are working in TT gauge, so that we reproduce the same results of Ref. \cite{Kehagias:2025xzm}.

\subsection{Generalization to higher order}
We can take a  step forward and  demonstrate that the non-linear behavior $t^{-2\ell-1}$ is a general solution that can be extended up to all the perturbative orders.
Let us  imagine, as an example, the Feynman diagram that describes the third perturbative order: in a vertex $(t_1,r_1)$ we have the scattering between two gravitons, that will produce a new third graviton. Now this graviton will scatter with another free graviton in a vertex $(t_3,r_3)$ (the one that we have studied before). The resulting new graviton will be contracted with the external field, becoming the observed tail.

Obviously, we need to attach to the external gravitons legs the usual classical legs, in order to avoid the formation of the tadpole.
The diagram under consideration is described by the following integral
\begin{eqnarray}\label{hfull}
    \langle h \rangle&=& \int_{t_0}^{t}dt_5\int_{t_0}^{t_5}dt_4\int_{t_0}^{t_4}dt_3\int_{t_0}^{t_3}dt_2\int_{t_0}^{t_2}dt_1\Big<\Big[H_{\,\,{\rm int}}^{{\rm q}\,{\rm cl}\,{\rm cl}}(t_5),\Big[H_{\,\,{\rm int}}^{{\rm q}\,{\rm cl}\,{\rm cl}}(t_4),\Big[H_{\,\,{\rm int}}^{{\rm q}\,{\rm cl}\,{\rm cl}}(t_2),\Big[H_{\,\,{\rm int}}^{{\rm q}\,{\rm q}\,{\rm q}}(t_1),\nonumber \\
    &&\left[H_{\,\,{\rm int}}^{{\rm q}\,{\rm q}\,{\rm q}}(t_3),h(t)\right]\Big]\Big]\Big]\Big]\Big>.
\end{eqnarray}
The first inner commutator is the same of the previous diagram in Eq. (\ref{Acommutator}) 
\begin{eqnarray}
    A&=&\left[H_{\,\,{\rm int}}^{{\rm q}\,{\rm q}\,{\rm q}}(t_3),h(t)\right]\simeq\int^{\infty}dr_3\; r_3^2 \; \;\int_{0}^{\infty}d\omega_1 \omega_1^5\omega'^{2}e^{-i\omega'(t_3-t)}\;j_{\ell_1}(\omega_1r)j_{\ell_2}(\omega_1r)j_{\ell_3}(\omega'r)\nonumber\\
    &\cdot& j_{\ell_3}(\omega'y)a_{\omega_1}a_{\omega_1}^{\dagger}.
\end{eqnarray}
Now we write again the usual Hamiltonian for a vertex with three gravitons
\begin{equation}
    H_{\,\,{\rm int}}^{{\rm q}\,{\rm q}\,{\rm q}}(t)\simeq\int^{\infty}dr\; r^2 \;\int_0^{\infty} d\omega_1\;\omega_1^5\int_0^{\infty}\frac{d\omega_3}{\sqrt{2\omega_3}}\omega_3^2
    e^{-i\omega_3t}\;j_{\ell_1}(\omega_1r)j_{\ell_2}(\omega_1r)j_{\ell_3}(\omega_3 r)a_{\omega_1}a_{\omega_1}^{\dagger}a_{\omega_3}.
\end{equation}
This is the generic term that we need to attach to the inner commutator to compose the final vacuum expectation value. Now we notice that the only component that provide information on the non linear tail (so the only integral that contain information on the frequency $\omega'$ that we send to zero by definition of the observed tail) is the inner commutator. All the other possible external commutator (related to vertices with respect to cubic graviton interaction or cubic vertex associated to external classical leg) will not give any contribution to the tail behavior, just for the momentum conservation and the angular momentum conservation applied to all the vertices of the diagrams. We can see this just by looking to the inner commutator: this does not  have any annihilation or creation  operator related to the external frequency $\omega'$.
Therefore we can argue that, at each perturbative order,  
\begin{equation}
    \langle h(t)\rangle \sim \frac{1}{t^{2\ell+1}}.
\end{equation}
This conclusion holds beyond the cubic order. Since the full  interaction Hamiltonian has always two derivatives in the gravitational field  in each   vertex order, the power of the  frequency of the final state will not change and  the power-law tail will not be modified.

There is another way to think of this  result. Imagine we write the equation of motion of a graviton with a source written, say, at the 27th order. Such a source will be composed by terms of the form $(\textrm{first order})^{27}$, $(\textrm{first-order})^{25}\cdot (\textrm{second-order}),\cdots, (\textrm{first-order})\cdot (\textrm{26th-order})$. Now, far from the source even the graviton calculated at the 26th-order will be a free propagating wave, like its first-order counterpart,  with an amplitude decaying like $1/r$. From the point of  view of the source, therefore, there will be always a term, at any order, decaying like $1/r^2$, the same which happens for the  second-order  source. We expect therefore the same power-law for the tail one obtains at second-order, but obviously with a much smaller overall amplitude.

\section{Conclusions}
Tails are a fundamental feature of the ringdown phase in gravitational wave signals from black holes — two cornerstones of general relativity. By employing the in-in formalism, we have rederived Price’s law for the linear tail and provided a transparent explanation for the power-law behavior of the nonlinear tail. Our interpretation rests on a simple picture: the generated QNMs scatter off the external classical field produced by the black hole itself in the asymptotic region.

There remain several avenues for refinement. One is to consider that a given nonlinear mode may arise from interactions with other nonlinear QNMs, whose dynamics could modify the power-law, even though their amplitudes are necessarily subleading \cite{Ling:2025wfv}. Another important question concerns the effects of interactions beyond the cubic level — in particular, whether the interaction with the external field can be resummed. We leave these questions for future work.

\section*{Acknowledgements}
We thank F. Bernardo, A. Kehagias and D. Perrone for many useful discussions.
L.L.B. thanks the Erasmus program for financial support and the University of Geneva for the kind hospitality during the realization of this project. 
A.I. acknowledges support from the  Swiss National Science Foundation (project number CRSII5\_213497). A.R. acknowledges support from the Swiss National Science Foundation (project number CRSII5 213497) and from the Boninchi Foundation for the project “PBHs in the Era of GW Astronomy”. 
\appendix
\section{Calculation of \texorpdfstring{$\mathcal{L}_{\rm tens}$}{L	extsubscript{tens}}}\label{AppendixA}
We consider the following setup. A graviton that propagates along the $x$-axis with momentum $k_1^{\,\mu}$ scatters off  a second graviton that is propagating along the $y$-axis, having momentum $k_2^{\,\mu}$. The collision between the two will originate a new graviton that will propagate along the diagonal direction, with a momentum $k_3^{\,\mu}$ given by the momentum conservation at the vertex of the interaction. The  three four-momenta in cartesian coordinates are therefore 
\begin{equation}
    \begin{split}
        &k_1^{\,\mu}=(\omega_1,\omega_1,0,0)\\
        &k_2^{\,\mu}=(\omega_2,0,\omega_2,0)\\
        &k_3^{\,\mu}=(\omega_3,\omega_3,\omega_3,0).
    \end{split}
\end{equation}
where we notice that the conjugate momenta from the interaction $\pi_{\alpha\beta}=\delta\mathcal{L}^{(3)}/\delta(\partial_0h^{\alpha\beta})$ are vanishing in the  TT gauge. In polar coordinates the polarization tensors along the $y$- and $d$-axis (identified by the momentum conservation)  are

\begin{equation}
\resizebox{\textwidth}{!}{%
$\displaystyle
    \epsilon^+_{\mu\nu,y}=\begin{pmatrix}
0 & 0 & 0 & 0 \\
0 & -\cos^2\theta + \cos^2\phi \sin^2\theta & r\cos\theta  \sin\theta + r\cos\theta \cos^2\phi  \sin\theta & -r\cos^2\phi  \sin^2\theta \sin\phi \\
0 & r\cos\theta  \sin\theta + r\cos\theta \cos^2\phi  \sin\theta & r^2\cos^2\theta \cos^2\phi  - r^2 \sin^2\theta  & -r^2\cos\theta \cos\phi  \sin\phi \sin\theta \\
0 & -r\cos\phi  \sin^2\theta \sin\phi & -r^2\cos\theta \cos\phi  \sin\phi \sin\theta & r^2 \sin^2\theta \sin^2\phi
\end{pmatrix}
$,
}
\end{equation}
\begin{equation}
\resizebox{\textwidth}{!}{%
$\displaystyle
\epsilon^\times_{\mu\nu,y}=\begin{pmatrix}
0 & 0 & 0 & 0 \\
0 & 2 \cos\theta \cos\phi \sin\theta & r\cos^2\theta\cos\phi - r\cos\phi \sin^2\theta & -r\cos\theta \sin\theta \sin\phi \\
0 & r\cos^2\theta \cos\phi - r\cos\theta \sin^2\theta & -2 r^2\cos\theta \cos\phi \sin\theta & r^2 \sin^2\theta \sin\phi \\
0 & -r\cos\theta \sin\theta \sin\phi & r^2 \sin^2\theta \sin\phi & 0
\end{pmatrix}$,
}
\end{equation}
\begin{equation}
\resizebox{\textwidth}{!}{%
$\displaystyle
    \epsilon^+_{\mu\nu,d}=\begin{pmatrix}
0 & 0 & 0 & 0 \\
0 & \frac{1}{4}(-1 - 3 \cos2\theta - 2 \sin^2\theta \sin2\phi) & -\frac{r}{4}\sin2\theta(-3 + \sin2\phi) & -\frac{r}{2}\cos2\phi \sin^2\theta \\
0 & -\frac{r}{4} \sin2\theta(-3 + \sin2\phi) & \frac{r^2}{4} (-1 + 3 \cos2\theta - 2 \cos^2\theta \sin2\phi) & -\frac{ r^2}{4} \cos2\phi \sin2\theta \\
0 & -\frac{r}{2} \cos2\phi \sin^2\theta & -\frac{r^2}{4} \cos2\phi \sin2\theta & \frac{r^2}{2} \sin^2\theta (1 + \sin2\phi)
\end{pmatrix}$,
}
\end{equation}
\begin{equation}
\resizebox{\textwidth}{!}{%
$\displaystyle
    \epsilon^\times_{\mu\nu,d}=\begin{pmatrix}
0 & 0 & 0 & 0 \\
0 & \sqrt{2} \cos\theta \sin\theta(\cos\phi - \sin\phi) & \frac{r\cos2\theta (\cos\phi - \sin\phi)}{\sqrt{2}} & -\frac{r\cos\theta \sin\theta(\cos\phi + \sin\phi)}{\sqrt{2}} \\
0 & \frac{r\cos2\theta (\cos\phi - \sin\phi)}{\sqrt{2}} & \sqrt{2} r^2\cos\theta  \sin\theta(\cos\phi + \sin\phi) & \frac{r^2 \sin^2\theta (\cos\phi + \sin\phi)}{\sqrt{2}} \\
0 & -\frac{r\cos\theta \sin\theta(\cos\phi + \sin\phi)}{\sqrt{2}} & \frac{r^2 \sin^2\theta (\cos\phi + \sin\phi)}{\sqrt{2}} & 0
\end{pmatrix}$.
}
\end{equation}
We can now write the momentum vectors in polar coordinates 
\begin{equation}
        k_1^{\,\mu}=\omega_1\left(1,\cos\phi \sin\theta,\frac{\cos\theta\cos\phi}{r},-\frac{\csc\theta \sin\phi}{r}\right),
\end{equation}
and
\begin{equation}
        k_2^{\,\mu}=\omega_2\left(1,\sin\phi \sin\theta,\frac{\cos\theta\sin\phi}{r},-\frac{\csc\theta\cos\phi}{r}\right),
\end{equation}
and
\begin{equation}
        k_3^{\,\mu}=\omega_3\left(2,(\sin\phi+\cos\phi) \sin\theta,\frac{\cos\theta(\sin\phi+\cos\phi)}{r},\frac{\csc\theta(\cos\phi-\sin\phi)}{r}\right).
\end{equation}
We obtain 
\begin{eqnarray}\label{LtensA}
 \mathcal{L}_{\rm tens}&=&\sqrt{G}\;\Big[ \frac{1}{4} \epsilon_1^{\,\alpha \beta} k_{2,\,\alpha} \epsilon_{2}^{\,\mu \nu} k_{3,\,\beta} \epsilon_{3\,\mu \nu} - \epsilon_{1}^{\,\mu \nu} k_{2,\,\alpha} \epsilon_{2,\,\mu}^{\,\alpha} k_{3,\,\beta} \epsilon_{3,\,\nu}^{\,\beta}
- \epsilon_1^{\,\mu \nu} k_{2,\,\nu} \epsilon_{2,\,\mu}^{\alpha} k_{3,\,\beta} \epsilon_{3,\,\alpha}^{\beta}\\
&+& \frac{1}{2} \epsilon_1^{\,\mu \nu} k_2^{\,\alpha} \epsilon_{2,\,\mu \nu} k_{3,\,\beta} \epsilon_{3,\,\alpha}^{\,\beta}\nonumber
+ \frac{1}{2} \epsilon_1^{\,\mu \nu} k_{2,\,\alpha} \epsilon_{2,\,\nu \beta} k_3^{\,\beta} \epsilon_{3,\,\mu}^{\,\alpha}+ \frac{1}{2} \epsilon_1^{\,\mu \nu} k_{2,\,\beta} \epsilon_{2,\,\nu \alpha} k_3^{\,\beta} \epsilon_{3,\,\mu}^{\,\alpha}+{\rm perm}(1,2,3)\Big]\nonumber \\= &&\frac{3\omega_1\omega_2 +3\omega_1\omega_3+3\omega_2\omega_3}{4}.      
\end{eqnarray}
In our case, we always consider $\omega_1=\omega_2$, and once the contraction with the external field is accounted for, we set $\omega_3\to 0$.
In such a  case we have
\begin{eqnarray}
     \mathcal{L}_{\rm tens}\simeq \frac{3\omega_1^2}{4}.
\end{eqnarray}
\section{One graviton and two classical source vertices}
\label{sec:PriceLaw}
In this section we want to provide the calculation of the cubic vertex with one quantized graviton that scatters two time with the background perturbation metric, far away from the BH.
We can write the cubic coupling in the following way
\begin{eqnarray}\label{5.1}
S_3[h] & = &4\pi\sqrt{G}\;\int dt \;dr\;r^2\;d\Omega \;
\Big[ \frac{1}{4} h_1^{\,\alpha \beta} \partial_\alpha h_2^{\,\mu \nu} \partial_\beta h_{3,\,\mu \nu} -  \, h_1^{\,\mu \nu} \partial_\alpha h_{2,\,\mu}^{\,\alpha} \partial_\beta h_{3,\nu}^{\beta}
- h_1^{\,\mu \nu} \partial_\nu  \, h_{2,\,\mu}^{\,\alpha} \partial_\beta h_{3,\,\alpha}^{\,\beta}\nonumber \\
& +& \frac{1}{2} h_1^{\,\mu \nu} \partial^\alpha h_{2,\,\mu \nu} \, \partial_\beta h_{3,\,\alpha}^{\,\beta}
+ \frac{1}{2} h_1^{\,\mu \nu} \partial_\alpha h_{2,\,\nu \beta} \, \partial^\beta h_{3,\,\mu}^{\,\alpha}\nonumber \\
&+& \frac{1}{2} h_1^{\,\mu \nu} \partial_\beta h_{2,\,\nu \alpha} \, \partial^\beta h_{3,\,\mu}^{\,\alpha}+{\rm{perms}}(1,2,3)
\Big],
\end{eqnarray}
in which we consider just as a quantized graviton $h_1$, and $h_2$, $h_3$ as classical sources. Those fields will, indeed, incorporate the information that we are not in a completely flat region of spacetime, but we are in presence of a BH.\\
We can extract the first order metric perturbation as
\begin{equation}
    h_{2,\,\mu\nu}= h_{3,\,\mu\nu}=\begin{pmatrix}
        2M/r&0&0&0\\
        0&2M/r&0&0\\
        0&0&0&0\\
        0&0&0&0
    \end{pmatrix}.
\end{equation}
Doing the contraction, as we did for the linear Price case, we get (for the $+$ polarization)
\begin{eqnarray}
\mathcal{L}_{\rm tens}&\simeq\;&\frac{5 M^2 \omega \left( 3 \cos3\phi \sin\theta 
+ \cos\phi \left( \sin\theta - 2(-3 + \cos2\phi) \sin3\theta \right) \right)}
{2 r^3} \nonumber\\
&+&\frac{24 M^2 \left( \cos^2\theta - \sin^2\theta \, \sin^2\phi \right)}{r^4}.
\end{eqnarray}
Now since we are working far away from the BH, so $r\gg2M$ we can just take the first leading order in $r$ (since the frequency $\omega$ will not be contracted with the external field it will be non zero). We can finally write the Hamiltonian of interaction (for the $+$ polarization)
\begin{eqnarray}
H_{\,\,{\rm int}}^{{\rm q}\,{\rm cl}\,{\rm cl}}&\simeq&\;4\pi\sqrt{G}\int\;dr\,r^2\,d\Omega\,\Bigg[\int\frac{d\omega\;\omega^2}{(2\pi)^{3/2}\sqrt{2\omega}}\sum_{\ell,m}i^\ell j_\ell(\omega r)Y_{\ell m}^*(\theta,\phi)e^{-i\omega t }\nonumber \\
&\cdot&\left[a^+(\omega)\frac{M^2\omega}{r^3}f(\theta,\phi)\right]+\rm{h.c}\Bigg]
\label{HintNLP}
\end{eqnarray}
where 
\begin{equation}
    f(\theta,\phi)\simeq \frac{5  \left( 3 \cos3\phi \sin\theta 
+ \cos\phi \left( \sin\theta - 2(-3 + \cos2\phi) \sin3\theta \right) \right)}
{2}.
\end{equation}
\bibliographystyle{JHEP}
\bibliography{draft}

\end{document}